\documentclass[aps,prb,reprint,superscriptaddress]{revtex4-2}
\usepackage{amsmath}
\usepackage{amssymb}
\usepackage{graphicx}
\usepackage{hyperref}
\usepackage{url}
\usepackage{epstopdf}
\usepackage{color,xcolor}

\begin{document}
\title{Effect of Rashba spin-orbit coupling on topological phases in monolayer ZnIn$_{2}$Te$_{4}$}
\author{Jun-Jie Zhang}
\email{junjiezhang@seu.edu.cn}
\affiliation{Key Laboratory of Quantum Materials and Devices of Ministry of Education, School of Physics, Southeast University, Nanjing 211189, China}
\affiliation{Department of Materials Science and NanoEngineering, Rice University, Houston, Texas 77005, USA}
\author{Shuai Dong}
\affiliation{Key Laboratory of Quantum Materials and Devices of Ministry of Education, School of Physics, Southeast University, Nanjing 211189, China}

\begin{abstract}
The interplay of Rashba and quantum spin Hall effects in non-centrosymmetric systems presents both challenges and opportunities for spintronic applications. While Rashba spin-orbit coupling can disrupt the quantum spin Hall phase, their coexistence may enable additional spintronic functionalities by coupling spin-momentum locking in the bulk with topologically protected edge states. Using the ZnIn$_2$Te$_4$ monolayer as a case study—a predicted polar two-dimensional topological insulator—we investigate how intrinsic and Rashba spin-orbit coupling compete within a single material. Our results identify key conditions under which sizable Rashba spin splitting can coexist with a stable quantum spin Hall phase, offering guidance for engineering quantum spin Hall insulators with enhanced spintronic capabilities.
\end{abstract}
\maketitle

\section{INTRODUCTIONS}
The realization of quantum spin Hall (QSH) insulator state in HgTe quantum wells has promoted intense interest in searching for nontrivial topological materials \cite{bernevig2006quantum,qi2010quantum}. Topological insulator (TI), a new classification of quantum materials, is characterized by conventional insulating bulk states and gapless edge states \cite{hasan2010colloquium}. One of the unique properties of TI is time-reversal symmetry protected edge states: spin of the edge state is locked at a right angle to the carrier momentum, making it very robust against local non-magnetic perturbations \cite{hasan2010colloquium}. To understand the origin of novel phenomena, comprehensive theoretical models \cite{kane2005quantum,bernevig2006quantum}, e.g., Kane-Mele-Hubbard (KMH) model \cite{kane2005quantum}, were constructed in past few decades. In these models, spin-orbit coupling (SOC) term plays an essential role in realizing QSH and inverting topological order in topological materials, i.e., topological trivial to non-trivial states. Based on the crystal symmetry, SOC can be divided into two categories: symmetry-independent and symmetry-dependent SOC \cite{manchon2015new}. The former, referred to as intrinsic SOC, arises from the spin-orbit interaction within atomic orbitals. The latter results from the coupling of the spin-orbit interaction to the crystal potential, due to the breaking of inversion symmetry.

Rashba SOC, a form of symmetry-dependent SOC, originates from either an external electric field or the presence of a non-centrosymmetric interface \cite{bihlmayer2015focus,liang2024ferroelectric}. Interestingly, Rashba SOC can give rise to additional unique quantum properties in TIs, such as crystalline-surface-dependent topological states, natural topological p-n junctions, and spin field-effect transistors \cite{wan2011topological,wang2012topological,feng2024manipulating}. However, according to the KMH model, Rashba SOC competes with intrinsic SOC, and can destroy the non-trivial topological bands if sufficiently strong \cite{laubach2014rashba,du2020competition}. This sensitivity is particularly evident in TIs, which are highly responsive to the experimental substrate \cite{kane2005quantum,zhou2014epitaxial}. Actually, these detrimental effects of Rashba SOC widely exist in polar 2D TI and even in ferroelectric TI heterobilayers \cite{zhang2020heterobilayer,tian2024ferroelectrically}. Hence, it is necessary to further investigate the Rashba effects on these systems.

In this work, inspired by experimental ZnIn$_{2}$S$_{4}$ \cite{lopez2001determination,wang2018construction,zhang2018mos2}, the non-centrosymmetric ZnIn$_{2}$Te$_{4}$ monolayer is predicted to be a promising candidate with the coexistence of Rashba SOC and mon-trivial band topology. The density functional theory (DFT) calculations and tight-binding (TB) model show that band gap is closed and reopened by strong intrinsic SOC, resulting in transition from trivial to non-trivial topological phases in ZnIn$_{2}$Te$_{4}$ monolayer. Furthermore, the topological phase is demonstrated by spin Chern number and topological edge state. Finally, the competition between intrinsic and Rashba SOC in ZnIn$_{2}$Te$_{4}$ monolayer is identified based on DFT calculations and then explained by TB model. The understanding of the competition between intrinsic and Rashba SOC may provide a new approach to designing giant Rashba in 2D polar or ferroelectric TI, making them attractive for applications in nanoscale spintronic devices.

\section{COMPUTATIONAL METHODS}
All calculations are performed using the plane wave basis Vienna $ab$ $initio$ simulation package (VASP) code \cite{kresse1996efficient,kresse1999ultrasoft}. The generalized gradient approximation in the Perdew-Burke-Ernzerhof (GGA-PBE) formulation is used with a cutoff energy of $450$ eV. The vacuum space of $\sim25$ {\AA} is intercalated into interlamination to eliminate the interaction between layers. A $12 \times 12$ 2D grid uniform $k$-points is applied for DFT calculations. The van der Waals (vdW) corrections were employed (DFT-D3) to calculate the interlayer distance in bulk phases \cite{grimme2010consistent}. The hopping energies of the TB model are calculated using Wannier representations \cite{mostofi2008wannier90}, where the Bloch states of Te-$p$ and Zn-$s$ orbitals are projected from DFT band structures. The six-band TB model using the basis of In1-$s\uparrow$, Te1-$p_{x}\uparrow$, Te1-$p_{y}\uparrow$, In1-$s\downarrow$, Te1-$p_{x}\downarrow$ and Te1-$p_{y}\downarrow$ is constructed to capture the ZnIn$_2$Te$_4$ monolayer orbital characteristics near the Fermi energy, while two-band effective TB model is employed to identify its spin-up/-down band topology.

\section{RESULTS AND DISCUSSIONS}
\begin{figure}
\centering
\includegraphics[width=0.45\textwidth]{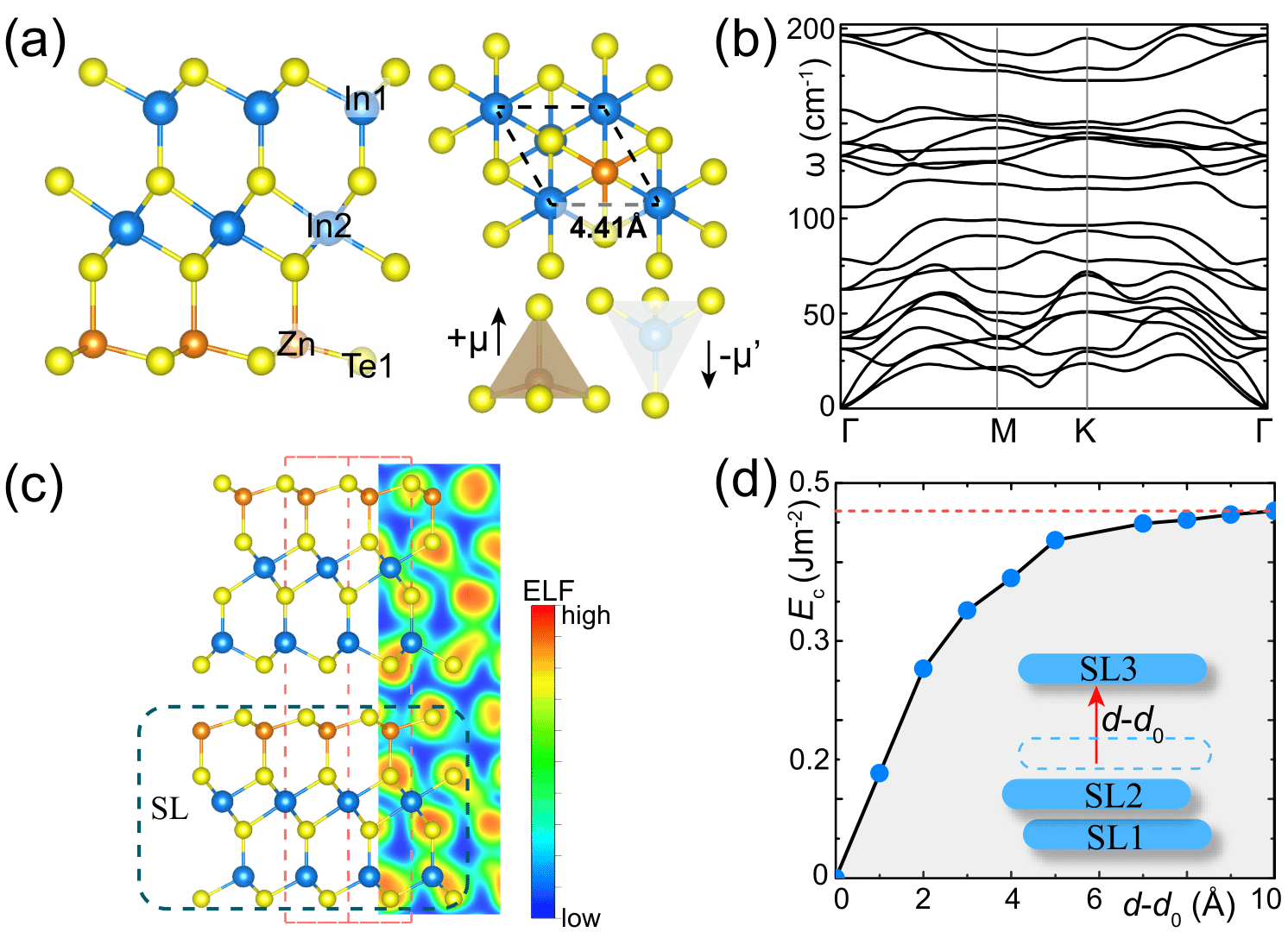}
\caption{(a) Top and side views of the ZnIn$_{2}$Te$_{4}$ monolayer. Insert: In-central and Zn-central tetrahedron. (b) Phonon dispersion of the ZnIn$_{2}$Te$_{4}$ monolayer. (c) Side views of the bulk ZnIn$_{2}$Te$_{4}$. On the right: calculated electron localization function. (d) Calculated exfoliation energy as function of separation distance ($d-d_{0}$), where $d_{0}$ indicates the vdW gap between adjacent layers in bulk phase.}
\label{Fig1}
\end{figure}

Inspired by experimental ZnIn$_{2}$S$_{4}$ \cite{lopez2001determination,wang2018construction,zhang2018mos2}, ZnIn$_{2}$Te$_{4}$ monolayer is supposed to have a 2D hexagonal layer structure with the polar point group $C_{3v}$. It is composed of a Te-Zn-Te-In-Te-In-Te septuplet layer (SL), where each layer is arranged in a triangular sublattice. Each SL has two kinds of In atoms: half in the tetrahedral environment (In1), and remaining half in the octahedral environment (In2), while Zn atoms are tetrahedral with four neighboring Te atoms [as illustrated in Fig. \ref{Fig1}(a)]. To assess the energetic stability, the formation energy is calculated as follows, $E_{b}=E_{tot}-(E_{Zn}-2E_{In}-4E_{Te})$, where $E_{tot}$, $E_{Zn}$, $E_{In}$ and $E_{Te}$ are the total energy of ZnIn$_{2}$Te$_{4}$, bulk Zn, bulk In, and bulk Te, respectively. The obtained formation energy is about $-0.89$ eV/cell, indicating ZnIn$_{2}$Te$_{4}$ monolayer is thermodynamically stable. The dynamic stability of ZnIn$_{2}$Te$_{4}$ is also identified by phonon calculation, showing no negative frequency in the whole phonon dispersion [Fig. \ref{Fig1}(b)]. After full optimization, two types of tetrahedral bonding, In-central and Zn-central tetrahedron, are not geometrically equal to each other due to the different atomic radii between Zn and In [Fig. \ref{Fig1}(a)], leading to internal electronic dipole moment. Based on the standard berry phase calculation \cite{king1993theory,resta1993towards}, the obtained Born effective charges of Zn and In1 are $0.28$$\lvert e \rvert$ and $0.36$$\lvert e \rvert$, respectively, and the total dipole moment is $1.07$ D/f.u. in ZnIn$_{2}$Te$_{4}$ monolayer.

To explore the experiential possibility of ZnIn$_{2}$Te$_{4}$, the corresponding bulk phase is considered by stacking the SL along the $c$-axis. The various stacking sequences are calculated in consideration of polymorphs of ZnIn$_{2}$S$_{4}$ \cite{lopez2001determination,wang2018construction,lee2019revisiting}. Unlike ZnIn$_{2}$S$_{4}$, the more stable bulk structure consists of two stacked SLs within a unit cell, which has a space group $P6_{3}mc$ [Fig. \ref{Fig1}(c)]. In particular, two stacked layers are twisted by $60^\circ$, generating a non-centrosymmetric bulk structure. Moreover, the electron localization function [Fig. \ref{Fig1}(c)] together with large interlayer distance (2.94 {\AA}) confirm that two stacked SLs are separated by a van der Waals (vdW) gap, which allows a ZnIn$_{2}$Te$_{4}$ monolayer to be exfoliated from its bulk phase. The calculated cleavage energy is $~0.46$ Jm$^{-2}$ when separation distance ($d-d_{0}$) is increased to $10$ {\AA} in a 3-SL slab [Fig. \ref{Fig1}(d)], which is close to the experimental value of graphite (0.36 Jm$^{-2}$) \cite{zacharia2004interlayer}, suggesting the experimental feasibility.

\begin{figure}
\centering
\includegraphics[width=0.47\textwidth]{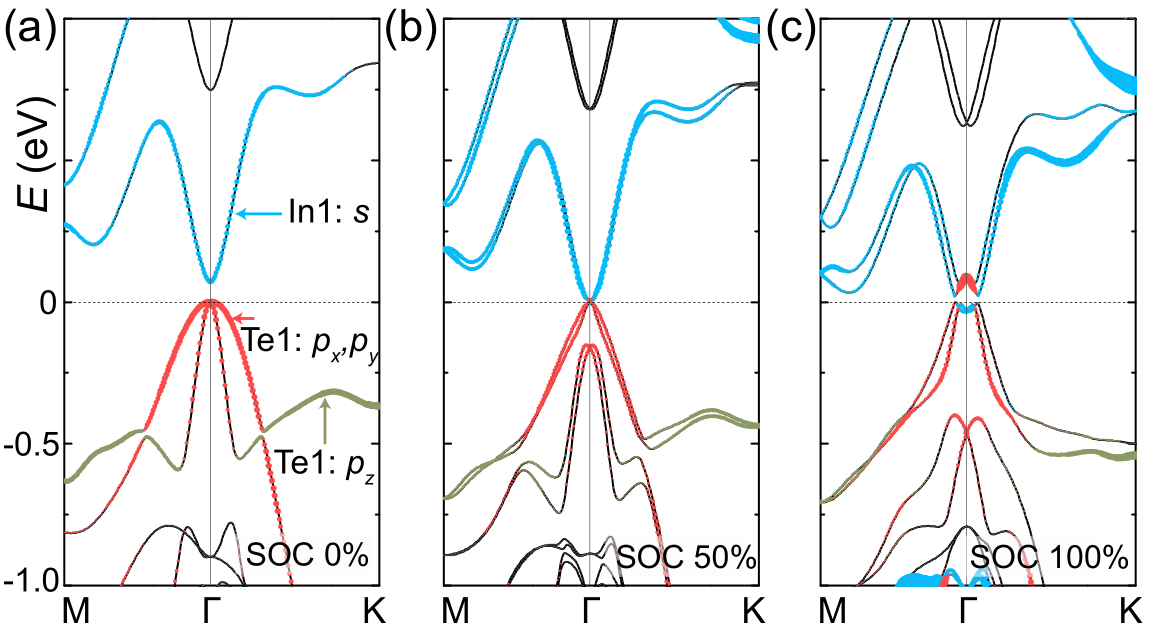}
\caption{Band structure around the Fermi level for ZnIn$_{2}$Te$_{4}$ monolayer with the inclusion of various SOC strength. The SOC strength ($\lambda$) is artificially modified by introducing a scaling factor $w$ such that the strength becomes $w\lambda$ in the DFT code. (a) $w = 0\%$; (b) $w = 50\%$; (c) $w = 100\%$. Blue color is the projection onto In1-$s$ orbitals, while red and yellow colors represent Te1-$p_x$,$p_y$, and $p_z$ orbitals, respectively.}
\label{Fig2}
\end{figure}

\begin{figure}
\centering
\includegraphics[width=0.43\textwidth]{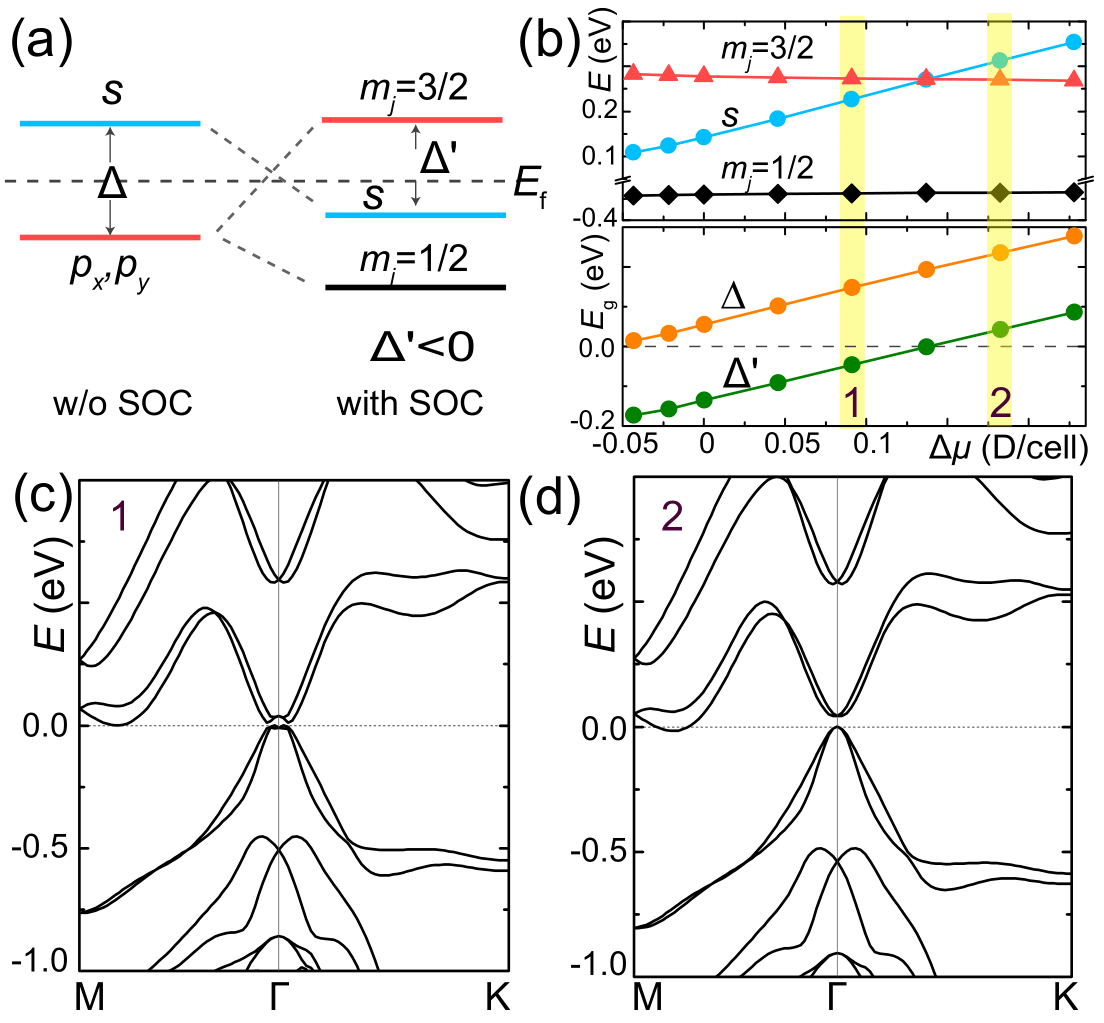}
\caption{(a) Schematics of the band evolution of ZnIn$_{2}$Te$_{4}$ monolayer at $\Gamma$ point. (b)Upper: the relative shift of energy levels ln1-$s$, Te1-$m_j$=3/2 and Te1-$m_j$=1/2 with the changes of the internal electronic dipole moment ($\Delta\mu = \mu - \mu_{0}$). The center of Te1-$m_j$=3/2 and Te1-$m_j$=1/2 state is defined as the energy zero. Lower: the changes of band gap at $\Gamma$ point without and with SOC. (c-d) band structures with various $\Delta\mu$ labeled in (b).}
\label{Fig3}
\end{figure}

We next focus on the electronic structures of ZnIn$_{2}$Te$_{4}$ monolayer. Both without and with inclusion of SOC band structures are shown in Fig. \ref{Fig2}. Without SOC, the valence band maximum (VBM) around $\Gamma$ point is derived mostly from outer Te1-$p_{x}$,$p_{y}$ orbitals, while the conduction band minimum (CBM) is predominantly In1-$s$ orbitals [Fig. \ref{Fig2}(a)]. Due to the higher energy of In1-$s$ orbitals than that of Te1-$p$ orbitals, the CBM and VBM are separated by a small direct band gap (about $55$ meV) at $\Gamma$ point. By turning on SOC, we observed that the band gap decreases to an inverted band point, around 50$\%$ of the total SOC strength [Fig. \ref{Fig2}(b)], and then it opens up again [about $23$ meV, Fig. \ref{Fig2}(c)]. In this process, the doubly degenerate Te1-$p_{x}$,$p_{y}$ orbitals are split into states: lower energy Te1-$m_{j}$=1/2 state and higher energy Te1-$m_{j}$=3/2 state [Fig. \ref{Fig3}(a)]. In particular, the strong SOC of Te atoms enables the band inversion between In1-$s$ and Te1-$m_{j}$=3/2 orbitals around $\Gamma$ point, as shown in Fig. \ref{Fig3}(a).

In contrast to centrosymmetric 2D systems, Rashba SOC exists in polar 2D structures commonly due to internal dipole moment, causing in-plane momentum-dependent splitting of electronic bands. Here, Rashba splitting is identified around $\Gamma$ point [Fig. \ref{Fig2}(d)], and then verified by the spin dispersion in Supplemental Material (SM) Fig. S1(a) \cite{sm} and spin texture in the $k_x - k_y$ plane of the two lowest conduction bands around $\Gamma$ point Fig. S2 \cite{sm}. The in-plane spin textures around the $\Gamma$ point exhibit identical spin-rotation directions for ``inner'' and ``outer'' states within the band inversion region (Fig. S2) \cite{sm}, deviating significantly from the conventional Rashba-like spin topology (see detailed discussion below). The helical spin texture outside the inversion region confirms the presence of the Rashba effect in the ZnIn$_2$Te$_4$ monolayer, resulting in a momentum-dependent spin splitting of these two lowest conduction bands. 

The Rashba constant $\alpha_{R}$ is calculated using the equation $2E_{R}/k_{R}$, where $E_{R}$ is Rashba splitting energy and $k_{R}$ is momentum offset as defined in Fig. S1(b) \cite{sm}. The obtained values of $\alpha_{R}$ are $1.41$, $0$ and $0.21$ eV{\AA} for Te1-$m_{j}$=1/2 state, VBM, and CBM, respectively. The value $\alpha_{R}$ of CBM in ZnIn$_2$Te$_4$ monolayer is somehow lower to that in In$_2$Si$_2$S$_3$Te$_3$ monolayer ($0.73$ eV{\AA}) \cite{mohanta20242}, but stronger than that in MoSSe monolayer ($66.1$ meV{\AA}) \cite{yu2021spin}. 

Besides the bands splitting, Rashba SOC can affect the band gap caused by intrinsic SOC in ZnIn$_{2}$Te$_{4}$ due to the $E_{R}$. It is noteworthy that the strength of Rashba SOC, associated with crystal potential and intrinsic SOC, can be turned by either internal dipole moment or external electric field. To illustrate the effects of Rashba SOC explicitly, internal dipole moment in ZnIn$_{2}$Te$_{4}$ is artificially changed by moving the atomic position of In1 or Te1, while the lattice constant $a_{0}$ remains constant. As shown in Fig. \ref{Fig3}(b), Rashba splitting around $\Gamma$ point becomes weaker with the decrease of dipole moment, leading to larger inverted energy gap between In1-$s$ and Te1-$m_{j}$=3/2 orbitals. In contrast, Te1-$m_{j}$=3/2 orbital shifts downward with respect to In1-$s$ orbital with the increase of dipole moment [Figs. \ref{Fig3}(b-d)]. These two orbitals are inverted at a critical dipole moment variation $\Delta \mu$ of around $0.14$ D/f.u. [Fig. \ref{Fig3}(b)], leading to the prevention of band inversion caused by intrinsic SOC [Fig. \ref{Fig3}(d)].

\begin{figure}
\centering
\includegraphics[width=0.48\textwidth]{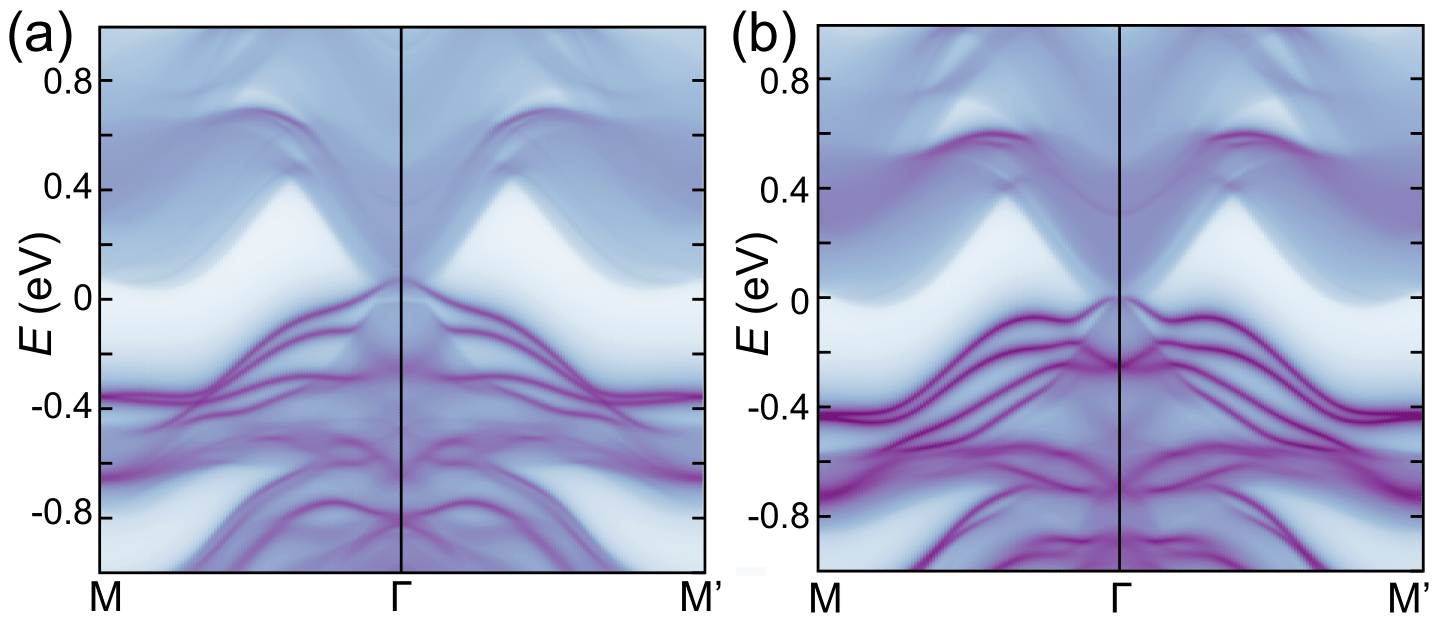}
\caption{Topological edge states along Zigzig direction on the boundary of  ZnIn$_{2}$Te$_{4}$ monolayer with (a) weak Rashba SOC ($\Delta\mu<0.14$ D/f.u.); (b) strong Rashba SOC ($\Delta\mu>0.14$ D/f.u.).}
\label{Fig4}
\end{figure}

Furthermore, we construct the edge Green's function of semi-infinite ZnIn$_{2}$Te$_{4}$ monolayer using the Wannier functions \cite{mostofi2008wannier90}. The calculated energy dispersion of the edge states with the inclusion of weak and strong Rashba SOC are shown in Fig. \ref{Fig4} and Fig. S4 \cite{sm}. With the weak Rashba SOC, the 2D bulk energy gap between conduction band and valence band is obviously closed by a pair of topological edge states, forming a single Dirac cone at $\Gamma$ point in Fig. \ref{Fig4}(a) and Fig. S4(a) \cite{sm}. However, this type of 1D gapless edge state is destroyed by strong Rashba SOC due to the lack of band inversion, leading to trivial insulating edge state [Fig. \ref{Fig4}(b) and Fig. S4(b)] \cite{sm}.

In order to further investigate the competition between intrinsic and Rashba SOC, the TB model is employed in the following discussion. Given the gapless edge state of  ZnIn$_{2}$Te$_{4}$ arises from the band inversion between In1-$s$ and Te1-$p_{x},p_{y}$ orbitals, a minimal basis of six orbitals (In1-$s\uparrow$, Te1-$p_{x}\uparrow$, Te1-$p_{y}\uparrow$, In1-$s\downarrow$, Te1-$p_{x}\downarrow$ and Te1-$p_{y}\downarrow$) is considered to construct the TB model. The corresponding TB Hamiltonian for ZnIn$_{2}$Te$_{4}$ monolayer is written as
\begin{equation}
\begin{split}
H{\rm{ = }}&\sum\limits_{\alpha ,i} {\sum\limits_\sigma  {{\varepsilon _{\alpha \sigma }}c_{i\alpha \sigma }^\dag {c_{i\alpha \sigma }}} }  + \sum\limits_{\left\langle {ij} \right\rangle \sigma } {\sum\limits_{\alpha ,\beta } {{t_{i\alpha \sigma ,j\beta \sigma }}c_{i\alpha \sigma }^\dag {c_{j\beta \sigma }}} }  + \\
&i{\lambda _{SO}}\sum\limits_{\left\langle {ij} \right\rangle \sigma ,\sigma '} {\sum\limits_{\sigma ,\sigma '} {c_{i\alpha \sigma }^\dag S_{\sigma \sigma '}^z{c_{j\beta \sigma '}}} }  +\\
&i{\lambda _R}\sum\limits_{\left\langle {ij} \right\rangle \sigma ,\sigma '} {\sum\limits_{\sigma ,\sigma '} {c_{i\alpha \sigma }^\dag {{\left( {{\bf{S}}_{\sigma \sigma '}^\mu  \times {\bf{d}}} \right)}^z}{c_{j\beta \sigma '}}} } \label{1}
\end{split}
\end{equation}
where the first two, third and last terms represent the Hamiltonian with onsite and nearest-neighbor hopping energy, intrinsic and Rashba SOC, respectively. The operator $c_{i\alpha \sigma }$ annihilates a particle with spin $\sigma$ of orbital $\alpha$ in $i$th cell, $\varepsilon_{\alpha}$ is onsite energy for orbital $\alpha$,and $t_{i\alpha \sigma,j\beta \sigma }$ is the nearest-neighbor hopping amplitude from $\beta$ orbital in $j$th cell to $\alpha$ orbital in $i$th cell ($\alpha$,$\beta$ indicate In1-$s$, Te1-$p_{x}$ and Te1-$p_{y}$). Here, the hopping between In1 and Te1 along $z$ direction is ignored, since they are arranged in different triangular sublattices and separated by the large distance along the out-of-plane [Fig. \ref{Fig1}(a)]. In the onsite intrinsic SOC term, $\lambda_{SO}$ is the intrinsic SOC strength. $S^{\mu}$ ($\mu$= $x$, $y$, $z$) denotes the three Pauli matrices, corresponding to the spin degree of freedom. Here, we considered only the on-site SOC term due to its highly localized interaction within each atom \cite{fernandez2007magnetic}. For the Rashba SOC, $\lambda_{R}$ is the Rashba SOC strength and vector $\bf{d}$ corresponds to the nearest-neighbor vectors $\bf{\delta_i}$ ($i$ = $x$,$y$,$z$). In the trigonal lattice of ZnIn$_2$Te$_4$ (${\bf{d}}=0$ in Eq.~\ref{1}), the Rashba SOC can effectively emerge through virtual hopping processes involving higher $p_z$ orbital (see detailed discussions in SM \cite{sm}). The Rashba SOC term breaks the spatial inversion symmetry, and couples spin-up and spin-down together for the orbitals of In1-$s$, Te1-$p_{x}$ and Te1-5$p_{y}$.

The total Hamiltonian matrix $H(k)$ is written in the basis of In1-$s\uparrow$, Te1-$p_{x}\uparrow$, Te1-$p_{y}\uparrow$, In1-$s\downarrow$, Te1-$p_{x}\downarrow$ and Te1-$p_{y}\downarrow$, as shown in Eq. S1 \cite{sm}. Confined by the symmetry of the system, there are eight independent parameters in total $H(k)$: $\varepsilon_s$, $\varepsilon_p$, $V_{ss\sigma}$, $V_{sp\sigma}$, and $V_{pp\sigma}$, $V_{pp\pi}$, $\lambda_{SO}$ and $\lambda_R$. $\varepsilon_s$ and $\varepsilon_p$ are on-site energies for $s$ and $p_x$,$p_y$ orbitals, respectively. $V_{ss\sigma}$, $V_{sp\sigma}$ and $V_{pp\sigma}$ are $\sigma$ type hopping energies, $V_{pp\pi}$ is $\pi$ type hopping energy. The on-site energies and hopping energies are obtained by Wannier function (Table \ref{Table1}), which fits well with DFT bands around $\Gamma$ point (Fig. S3) \cite{sm}.

\begin{table}
\centering
\caption{The obtained on-site energies and hopping energies based on Wannier function.}
\begin{tabular*}{0.4\textwidth}{@{\extracolsep{\fill}}lcccccc}
\hline
\hline
$~$    &$\varepsilon_s$   &$\varepsilon_p$   &$V_{ss\sigma}$  &$V_{sp\sigma}$  &$V_{pp\sigma}$  &$V_{pp\pi}$\\
\hline
E(eV)   &$3.06$  &$-2.46$    &$-0.50$  &$0.04$  &$0.02$ &$0.80$ \\
\hline
\hline
\end{tabular*}
\label{Table1}
\end{table}

\begin{figure}
\centering
\includegraphics[width=0.47\textwidth]{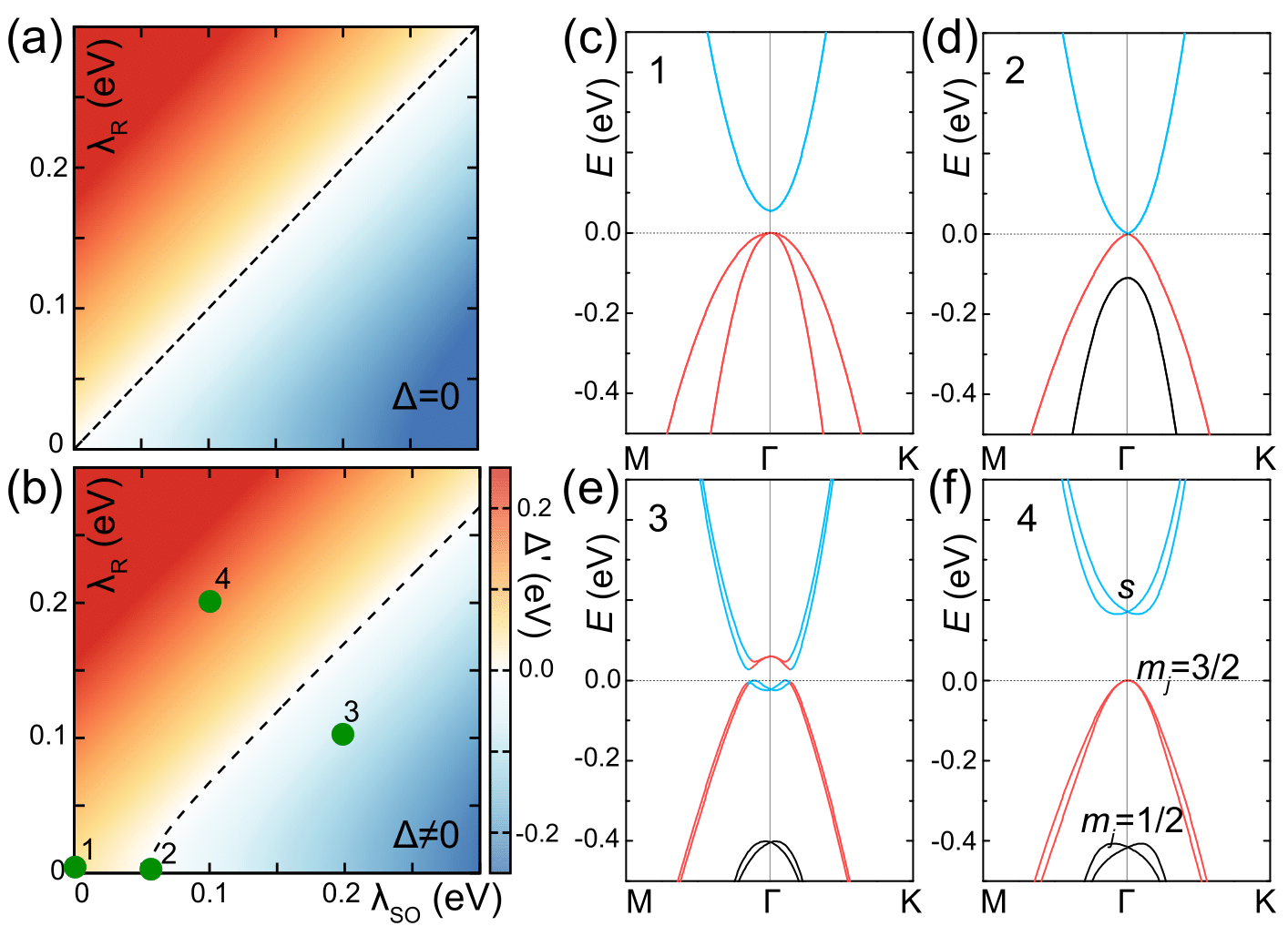}
\caption{Schematics of interacting $\lambda_R$-$\lambda_{SO}$ phase diagram at $\Gamma$ point: (a)$\Delta=0$ and (b) $\Delta=55$ meV. (c-f) Four different phases with $\lambda_{SO}$ and $\lambda_{R}$ labeled in (b). The onsite and hopping energies used in TB model are listed in Table \ref{Table1}.}
\label{Fig5}
\end{figure}

Without the inclusion of intrinsic and Rashba SOC, we obtained a non-degenerate eigenvalue ${E_s} = {\varepsilon _s} + 6{V_{ss\sigma }}$ and two degenerate eigenvalues $
{E_{{p_x},{p_y}}} = {\varepsilon _p} + (3{V_{pp\pi }} + 3{V_{pp\sigma }})$  at $\Gamma$ point, so the corresponding band gap is $\Delta  = {E_s} - {E_{{p_x},{p_y}}}$ [Fig. \ref{Fig3}(a)]. When two types of SOC are turned on, the eigenvalues at $\Gamma$ point are changed as follows,
\begin{equation}
\begin{split}
&\resizebox{.9\hsize}{!}{${E'_{s, \uparrow }} = {E'_{s, \downarrow }} = \frac{1}{2}\left( {{E_s} + {E_{{p_x},{p_y}}} - {\lambda _{SO}} + \sqrt {8\lambda _R^2 + {{\left( {\Delta  + {\lambda _{SO}}} \right)}^2}}}\right)$}\\
&{E_{{m_j} = 3/2, \uparrow }} = {E_{{m_j} = 3/2, \downarrow }}{\rm{ = }}{E_{{p_x},{p_y}}} + {\lambda _{SO}}\\
&\resizebox{.9\hsize}{!}{${E_{{m_j} = 1/2, \uparrow }} = {E_{{m_j} = 1/2, \downarrow }} = \frac{1}{2}\left( {{E_s} + {E_{{p_x},{p_y}}} - {\lambda _{SO}} - \sqrt {8\lambda _R^2 + {{\left( {\Delta  + {\lambda _{SO}}} \right)}^2}} } \right)$}\label{4}
\end{split}
\end{equation}
Among these orbitals, $s$ orbital and $m_j$=$3/2$ state occupy around Fermi level, while $m_j$=$1/2$ state has lower energy. Hence, the energy difference between $s$  orbital and $m_j$=$3/2$ state can be written as,
\begin{equation}
\begin{split}
\Delta '&{\rm{ = }}{E'_s} - {E_{{m_j} = 3/2}} \\
&= \frac{1}{2}\left( {\Delta  - 3{\lambda _{SO}} + \sqrt {8{{ \lambda _{R}^2}} + {{\left( {\Delta  + {\lambda _{SO}}} \right)}^2}} } \right).\label{22}
\end{split}
\end{equation}
Based on Eq. \ref{22}, the band gap decreases to an inverted band point when the value of $\Delta '$ becomes negative, and the obtained critical point is
\begin{equation}
{\lambda _R} < \sqrt {{\lambda _{SO}}\left( {{\lambda _{SO}} - \Delta } \right)} {\rm{  }}({\lambda _{SO}} > \Delta).\label{9}
\end{equation}

\begin{figure*}
\centering
\includegraphics[width=0.7\textwidth]{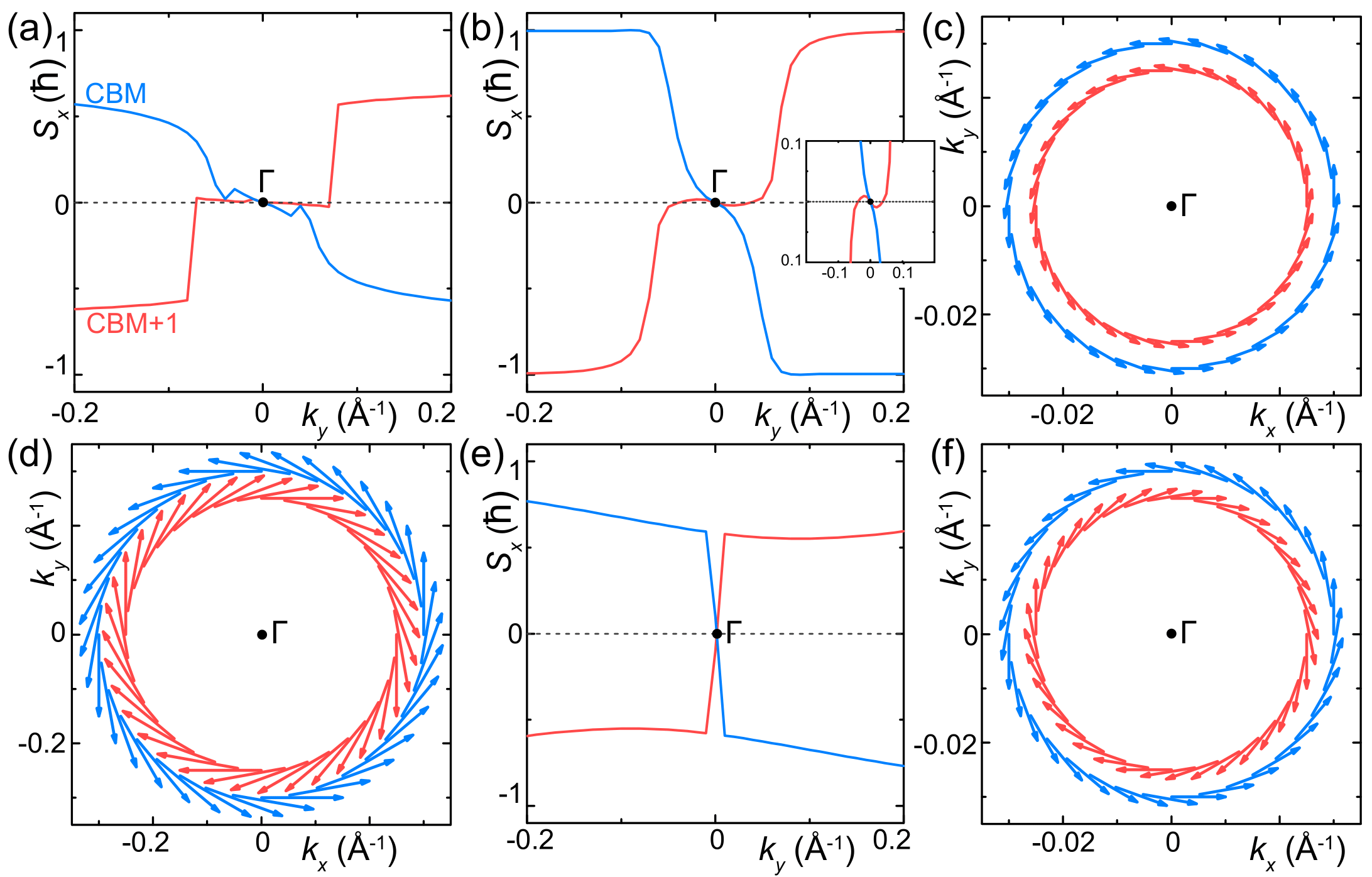}
\caption{(a-b) Calculated $S_x$ expectation along the $k_y$ direction (-$M$ – $\Gamma$ – $M$ path) for the two lowest conduction bands around $\Gamma$ point in phase 3 of ZnIn$_2$Te$_4$ monolayer: (a) DFT results and (b) TB results with $\lambda_{SO} = 0.2$ eV and $\lambda_{R} = 0.1$ eV. Inset: enlarged view of the nonlinear twisting behavior within band inversion region. (c-d) TB-calculated in-plane spin textures for phase 3: (d) within and (e) outside band inversion region. (e-f) TB-calculated spin topology around $\Gamma$ point for phase 4 with $\lambda_{SO} = 0.1$ eV, $\lambda_{R} = 0.2$ eV: (e) $S_x$ expectation; (f) in-plane spin textures.}
\label{Fig6}
\end{figure*}

As the final results, the interacting $\lambda_R$-$\lambda_{SO}$ phase diagram based on Eq. \ref{22} is shown in Figs. \ref{Fig5}(a-b). Obviously, to obtain nontrivial topological phases, the strength of Rashba SOC is required to be weaker than intrinsic SOC when $s$ and $p_{x}$,$p_{y}$ orbitals are degenerate [$\Delta=0$, Fig. \ref{Fig5}(a)] However, the situation becomes more complicate for $\Delta\neq0$, as shown in Fig. \ref{Fig5}(b). In details, the band gap will be closed when $\lambda_{SO}=\Delta$ without Rashba SOC [Fig. \ref{Fig5}(c)], and then it re-opens up ($\Delta '{\rm{ = }}{\lambda _{SO}} - \Delta $ at $\Gamma$ point) if intrinsic SOC is strong enough, which agrees well with DFT results (Fig. \ref{Fig2}). With the small Rashba SOC and strong intrinsic SOC, band inversion between $s$ orbital and $m_{j}$=3/2 state retains [Fig. \ref{Fig5}(d) and Fig. \ref{Fig3}(b)], while the band gap caused by intrinsic SOC is decreased (as illustrated by Eq. \ref{22}). However, the nontrivial topological state is driven into trivial state by the strong Rashba SOC [Fig. \ref{Fig5}(e)], which is also consistent with DFT results [Fig. \ref{Fig3}(b) and \ref{Fig3}(d)].

On the other hand, the band inversion around the $\Gamma$ point in the ZnIn$_2$Te$_4$ monolayer, driven by intrinsic SOC, is also associated with an unconventional spin topology in momentum space, where the spin-rotation directions of the ``inner'' and ``outer'' states are identical (Fig. S2). This unconventional spin configuration has also been predicted for certain $p_{x,y}$-derived states in BiTeI \cite{bawden2015hierarchical}, as well as in Bi/Cu(111) \cite{mirhosseini2009unconventional} and Pb/Cu(111) \cite{bihlmayer2007enhanced} surface alloys. The origin of this effect could be examined by analysis of spin expectation and textures using the TB model (Eq.~\ref{1}) \cite{szalowski2023spin}.

For phase 3, where band inversion occurs between the In1-$s$ and Te1-$p_{x,y}$ orbitals around $\Gamma$ point [Fig.~\ref{Fig5}(e)], both DFT- and TB-calculated $S_x$ expectation values of the two lowest conduction bands along the $k_y$ direction exhibit a nonlinear twisting behavior within band inversion area [Figs.~\ref{Fig6}(a-b)]. This twisting behavior reflects their identical spin rotation for ``inner'' and ``outer'' states [Fig.~\ref{Fig6}(c)]. However, outside the inversion region, the spin topology transitions to the conventional Rashba-type [Fig.~\ref{Fig6}(d)]. When band inversion is suppressed by the strong Rashba SOC, corresponding to phase 4 in Fig.~\ref{Fig5}(f), this nonlinear twisting of $S_x$ expectation disappears [Fig.~\ref{Fig6}(e)], resulting in a conventional Rashba-type spin configuration [Fig.~\ref{Fig6}(f)]. Hence, the origin of unconventional spin textures for two lowest conduction bands are associated with band inversion between the lower Te1-$p_{x,y}$ and higher In1-$s$ orbitals around the $\Gamma$ point.

Finally, the two-band effective TB model is used to study the topological properties of ZnIn$_{2}$Te$_{4}$, where spin-up/-down state is written as,
\begin{equation}
H = {\bf{R}}\left( {\bf{k}} \right)\cdot \bf{\sigma} = \left( {\begin{array}{*{20}{c}}
{{R_z}}&{{R_x} - i{R_y}}\\
{{R_x} + i{R_y}}&{ - {R_z}}
\end{array}} \right)\label{11}
\end{equation}
where $\bf{\sigma}$ are Pauli matrices, and $R_i$ ($i$ = $x$,$y$,$z$) are smooth functions of $k_x$ and $k_y$ with period $2\pi$. To obtain $R_i$ and keep $SU(2)$ symmetry, an effective SOC is considered in Eq. \ref{11}, which is defined as $\tilde \lambda = \frac{1}{2}\left(\Delta  + 3\lambda _{SO} - \sqrt{8 \lambda _{R}^2 + \left( \Delta  + \lambda _{SO} \right)^2} \right)$. In this formulation, the effective SOC term considers only the Rashba SOC strength, whereas the explicit Rashba terms are removed from Eq.~\ref{11}. Therefore, the basis of Eq. \ref{1} for spin-up/-down state is changed to ($s$, $p_x+ip_y$, $p_x-ip_y$) as shown in Eq. S3 \cite{sm}. Here, the low energy orbital ($p_x-ip_y$) is eliminated, and the smooth functions are written as
\begin{equation}
\begin{split}
&{R_x}{\rm{ = }}\frac{i}{{\sqrt 2 }}{h_{sx}}\\
&{R_y}{\rm{ = }}\frac{i}{{\sqrt 2 }}{h_{sy}}\\
&{R_z}{\rm{ = }}\frac{1}{2}\left[ {{h_{ss}} - \frac{1}{2}\left( {{h_{xx}} + {h_{yy}}} \right) - \tilde \lambda } \right]
\end{split}
\label{12}
\end{equation}
Therefore, the Berry curvature for the spin-up and spin-down states can be calculated independently, which is given by
\begin{equation}
{\Omega _{xy}} = {\bf{\hat R}}\left( \bf{k} \right)\frac{{\partial {\bf{\hat R}}}}{{\partial {k_x}}} \times \frac{{\partial {\bf{\hat R}}}}{{\partial {k_y}}},
 \label{14}
\end{equation}
where ${\bf{\hat R}}\left( \bf{k} \right){\rm{ = }}{{{\bf{R}}\left( \bf{k} \right)} \mathord{\left/
 {\vphantom {{{\bf{R}}\left( \bf{k} \right)} {\left| {{\bf{R}}\left( \bf{k} \right)} \right|}}} \right.
 \kern-\nulldelimiterspace} {\left| {{\bf{R}}\left( \bf{k} \right)} \right|}}$ is a unit vector. The corresponding Chern number for spin up (spin dn) is integrated over the entire Brillouin zone,
\begin{equation}
C = \frac{1}{{2\pi }}\int {{\Omega _{xy}}{d^2}k = } \frac{1}{{4\pi }}\int {{\bf{\hat R}}\left( \bf{k} \right)\frac{{\partial {\bf{\hat R}}}}{{\partial {k_x}}} \times \frac{{\partial {\bf{\hat R}}}}{{\partial {k_y}}}{d^2}k}.
\label{13}
\end{equation}

\begin{figure}
\centering
\includegraphics[width=0.47\textwidth]{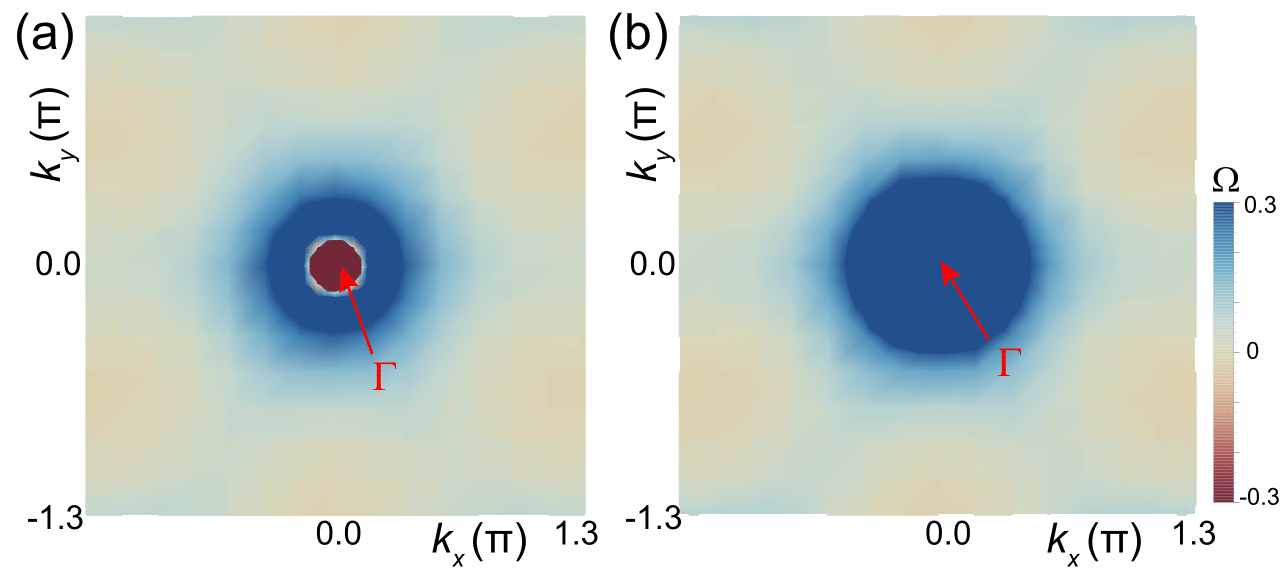}
\caption{The distribution of Berry curvature for spin-up state in the first Brillouin zone for (a) $\tilde \lambda<\Delta$ and (b) $\tilde \lambda>\Delta$. Note that the color schemes are identical in (a) and (b).}
\label{Fig7}
\end{figure}

The calculated Berry curvatures without and with band inversion are shown in Figs. \ref{Fig7}(a-b). When system has weak intrinsic SOC or strong Rashba ($\tilde \lambda  < \Delta $), the band inversion between $s$ and $m_j$=3/2 orbitals is absent, inducing the sign changes of Berry curvatures for spin-up state around $\Gamma$ point [Fig. \ref{Fig7}(a)]. In Fig. \ref{Fig7}(a), the positive background of Berry curvature cancels the negative contributions around $\Gamma$ point, giving a zero Chern number ($C$=$0$). With band inversion ($\tilde \lambda  > \Delta $), the Berry curvatures have peaks around $\Gamma$ point [Fig. \ref{Fig7}(b)], implying a non-zero Chern number $C$$\uparrow$=1 ($C$$ \downarrow$=-1) for spin up (spin down) sector. When time-reversal symmetry is present, the total Chern number is zero because time-reversal symmetry relates states with opposite momentum, effectively canceling out any topological contribution to the wavefunction. However, in this case, we used the spin Chern number instead, defined as $C_s = (C_{\uparrow} - C_{\downarrow})/2$ \cite{sheng2006quantum}, which is related to the $Z_2$ invariant. Hence, the non-zero spin Chern number ($C_s$=1) directly confirms a nontrivial topological phase in ZnIn$_2$Te$_4$ in case $\tilde \lambda>\Delta$. The effective SOC term captures the physical competition between intrinsic and Rashba SOC. The stronger $\lambda_R$ results in a weaker effective SOC strength. Although the explicit Rashba term is removed from Eqs.~\ref{11}, ~\ref{12}, ~\ref{14}, and ~\ref{13}, the effective SOC formulation reproduces the same band inversion region as obtained from Eq. S1 [Fig. \ref{Fig5}(b)], indicating an identical topological phase transition boundary (topological non-trivial state when $\lambda_R < \sqrt{\lambda_{SO}^2-\lambda_{SO}\Delta}$ and $\tilde \lambda>\Delta$).

To verify our TB model results, we recalculated the $Z_2$ invariant by analyzing the evolution of the Wannier charge center (WCC) \cite{taherinejad2014wannier}, as shown in Fig. S5 \cite{sm}, where all occupied bands throughout the Brillouin zone are considered. When WCC is crossed by any arbitrary horizontal reference lines an ``odd'' number of times, the whole system is topological nontrivial state, $Z_2 = 1$. (Detailed discussion of WCC method can be found in Ref. \cite{taherinejad2014wannier}). As expected, the WCC results indicate $Z_2=1$ ($Z_2=0$) for ZnIn$_2$Te$_4$ monolayer with weak (strong) Rashba SOC, which agrees well with our TB model findings.

\section{CONCLUSION}
In summary, we demonstrate the nontrivial topological phase in noncentrosymmetric ZnIn$_2$Te$_4$ monolayer. Considering the excellent dynamical and thermal stability, and the fact that its cleavage energy is very close to that of graphene, ZnIn$_2$Te$_4$ monolayer is a promising candidate to realize the coexistence of Rashba and QSH effects. The giant Rashba energy splitting is illustrated to be emerging from its internal dipole moment, due to geometrical non-equivalent between Zn-central and In-central tetrahedron. Based on DFT calculations and TB model, the competition between intrinsic and Rashba SOC in ZnIn$_2$Te$_4$ monolayer is identified, where strong Rashba effects will lead to topological transition from non-trivial to trivial state. The understanding of the competition between intrinsic and Rashba SOC may provide an interesting platform for exploring new physics and designing 2D polar TI with giant Rashba effects.

\acknowledgments{This work was supported by the National Natural Science Foundation of China (Grant No.~12404102), the Natural Science Foundation of the Jiangsu Province (Grant No.~BK20230806), open research fund of Key Laboratory of Quantum Materials and Devices of Ministry of Education (Southeast University), and Southeast University Interdisciplinary Research Program for Young Scholars (Grant No.~2024FGC1008). Most calculations were done on Big Data Center of Southeast University.}

\bibliography{ref}
\end{document}